\documentclass[14pt]{article}
\usepackage{graphicx,bm,color,amsmath,amssymb}
\usepackage{jheppub}

\newcommand{\ba}{\begin{eqnarray}}
\newcommand{\ea}{\end{eqnarray}}
\newcommand{\beq}{\begin{eqnarray}} 
\newcommand{\eeq}{\end{eqnarray}} 
\newcommand{\bea}{\begin{array}{rl}}
\newcommand{\eea}{\end{array}}
\newcommand{\ld}{\mathcal{L}}
\newcommand{\bel}{\begin{array}{l}}
\newcommand{\non}{\nonumber}
\newcommand{\D}{\nabla}
\def\bse{\begin{subequations}}
\def\ese{\end{subequations}}

\begin{document}

\title{Cosmological perturbations in generalised dark Lagrangians}
\author{James Edholm$^\dag$*,}
\author{Jonathan Pearson*}
\emailAdd{james.edholm@hotmail.co.uk}
\affiliation{${}^\dag$Consortium for Fundamental Physics, Physics Department,
Lancaster University, Bailrigg, Lancaster, LA1 4YW, U.K.}
\affiliation{*Centre for Particle Theory, Department of Mathematical Sciences,
Durham University, South Road, Durham, DH1 3LE, U.K.}
\date{\today}

\abstract{
We describe a new method to parameterise dark energy theories including massive gravity, elastic dark energy 
and tensor-metric theories. We first examine the existing framework which describes any second order Lagrangian which depends on the 
variation of the metric and find new constraints on the parameters. We extend the method to  Lorentz violating theories which depend on the 
variation of the time and spatial parts of the metric separately.
We show how this can describe massive gravity and elastic dark energy, while
ruling out the whole class of theories where the Lagrangian depends only on the
variation of the time part of the metric.

We further generalise our method to tensor-metric theories, both with and without splitting the metric into time and spatial parts.
Our method extends existing physics by providing a mechanism to easily evaluate large classes of dark energy theories.}
\maketitle

\section{Introduction}
The discovery of the accelerated expansion of the universe
\cite{Perlmutter:1998np,Riess:1998cb,Heymans:2012gg,Ade:2013kta,Ade:2013tyw,
Abbott:2005bi,Abate:2012za,Battye:2018ssx}
prompted many attempts to explain the phenomenon, caused by an unknown ``dark energy'', 
using modifications to Einstein's theory of General Relativity \cite{Clifton:2011jh}.
These theories generally started with a specific dark Lagrangian which was then investigated
to test its compatibility with both dark energy and other cosmological
and solar system observations. Here we follow the method of \cite{Pearson:2012kb,Bloomfield:2013cyf,Battye:2013er,Battye:2012eu,Battye:2013ida,Skordis:2008vt,Baker:2011jy,Clifton:2011jh,Baker:2012zs} and look at a \textit{class}
of dark Lagrangians, so that future experimental results can immediately
rule out a whole swathe of models without the 
need for further calculation.

We use the formalism introduced in \cite{Pearson:2012kb,Bloomfield:2013cyf,Battye:2013er,Battye:2012eu,Battye:2013ida}
which draws on the ``Post Parameterised Friedmann" (PPF) approach described in 
\cite{Skordis:2008vt,Baker:2011jy,Clifton:2011jh,Baker:2012zs}.
In these papers the Einstein-Hilbert action is modified with a \emph{dark
Lagrangian} such that the modified action is written as 
\beq \label{eq:modifiedeinsteinhilbert}
S = \int d^4 x \sqrt{-g} \left[\frac{R}{16 \pi G} - \ld_{matter} - \ld_d
\right],
\eeq
and equations of state for dark sector perturbations are found for the entropy
perturbation and the anisotropic stress.
The recent observation of a neutron star merger by LIGO \cite{TheLIGOScientific:2017qsa} 
severely constrained the speed of gravitational
waves, and seemingly placed strict restrictions on many modified gravity theories. 
However, it was suggested in \cite{deRham:2018red} that although these theories 
predict the speed of gravitational
waves to be different from that of light at low energies, this may not necessarily be the
case at the high energies seen in neutron star mergers, due to the unknown
 UV completions of these theories.

We find the perturbed fluid variables
when $\ld_{\{2\}} = \ld_{\{2\}}(\delta_L g_{\mu\nu})$,
in a similar way to \cite{Battye:2013er}. We then impose invariance under
time reparameterisation and find the equations of motion.
We also find evolution equations for the equation of state parameter $w$
and the elastic bulk modulus. 
We find restrictions on $w$ such that we can have a realistic (positive and
subluminal) sound speed. 
We find the perturbed fluid variables when we have imposed spatial invariance,
and we find the conditions under which the entropy is gauge invariant.

In this paper our main focus is on generalising all of these calculations for the
case where $\ld_{\{2\}}$ is a function of the change in the time-like and spatial parts
of the metric separately, 
not necessarily packaged together as the spacetime metric.

Elastic dark energy (EDE) discussed in \cite{Battye:2013er,Pearson:2014iaa} 
is a development of Carter-Quintana relativistic elasticity theory 
\cite{Carter:1982xm,Carter:1973zz,Carter:1980zz,Friedman:1975ab,
Carter:1972az}. These models describe the universe as a solid with certain
parameters which can be found by observations. 
By imposing time translational invariance but not spatial invariance, we
fulfill the conditions necessary for EDE. 

As we are adding a dark energy term to the Einstein-Hilbert action (\ref{eq:modifiedeinsteinhilbert}),
we can decompose the stress energy tensor as $T_{\mu\nu} = T_{\mu\nu}^{\text{matter}}
 + T_{\mu\nu}^{\text{radiation}}
+ T_{\mu\nu}^{\text{dark}}$, 
i.e. we have added a source term to the standard model stress energy tensor.

\subsection{Decomposition of the metric}

We impose spatial isotropy by foliating the four-dimensional (4D) space-time,
as in \cite{Battye:2013er,Tattersall:2017eav}, 
by three-dimensional (3D) sheets with a time-like unit vector, $u_\mu$ which
is everywhere orthogonal to the sheets. 
The 4D spacetime has metric $g_{\mu\nu}$, and the 3D sheets have spatial
metric $\gamma_{\mu\nu}
= \gamma_{(\mu\nu)}$. The (3 + 1) decomposition of the 4D metric is 
\beq
        g_{\mu\nu} = \gamma_{\mu\nu} - u_\mu u_\nu, 
\eeq
where $ u_\mu $ and $ \gamma_{\mu\nu} $ are subject to the orthogonality
and normality conditions:
\beq
        u^\mu\gamma_{\mu\nu} = 0,\qquad u^\mu u_\mu = -1.
\eeq
Hence we can write the dark term in the stress-energy tensor in an isotropic
spacetime as 
\beq \label{fluid-emt}
        T_{\mu\nu} = \rho u_\mu u_\nu + P \gamma_{\mu\nu},
\eeq
where $\rho$ is the density of the universe due to this dark term
and $P$ is the pressure. The pressure and density are related by $P = w \rho$,
where $w$ is the equation of state parameter. 
From now on, when we write $T_{\mu\nu}$, this will refer to the dark stress-energy
tensor rather than the overall energy momentum tensor.

\section{$\ld_{\{2\}}(\delta_L g_{\mu\nu})$} 
\label{sec:varfullmetriconly}
We will first look at the case $\ld_{\{2\}}(\delta_L g_{\mu\nu})$ before generalising 
to more generic Lagrangians. For any Lagrangian which is a function only of 
perturbations of the metric and no derivatives
thereof, i.e. $\ld _{\{2\}}= \ld_{\{2\}}(\delta_L g_{\mu\nu})$,
then from \cite{Battye:2013er}, the most general quadratic Lagrangian for
dark sector perturbations is 
\beq \label{eq:L}
        \ld_{\{2\}}= \frac{1}{8} W^{\mu\nu\alpha\beta} \delta_L g_{\mu\nu}
\delta_L g_{\alpha\beta},
\eeq
where $\delta_L g_{\mu\nu}$ is the metric fluctuation under a perturbation
and $W^{\mu\nu\alpha\beta}$ can be thought of as a 
mass term for the perturbation, not to be
confused with the Weyl tensor. From  \cite{Battye:2012eu}, 
we can use the symmetries of the $W$ tensor:
\beq 
        W_{\mu\nu\alpha\beta} = W_{(\mu\nu)(\alpha\beta)} = W_{\alpha\beta\mu\nu},
\eeq
to obtain the most general possible $W_{\alpha\beta\mu\nu}$: 
\beq \label{eq:1}
                W_{\mu\nu\alpha\beta} =  A_W u_{\mu}
u_{\nu} u_{\alpha} u_{\beta}  +B_W \left( u_{\mu} u_{\nu} \gamma_{\alpha\beta}
+ \gamma_{\mu\nu} u_{\alpha} u_{\beta} \right)
                + C_W u_{(\mu} \gamma_{\nu)(\alpha} u_{\beta)}  
                +  D_W \gamma_{\mu\nu} \gamma_{\alpha\beta} + E_W \gamma_{\mu(\alpha}
\gamma_{\beta)\nu}.~~~~~
\eeq
\subsection{Contractions of the Eulerian change in the energy-momentum tensor}
The deformation vector, which represents possible coordinate 
changes \cite{Battye:2013ida}\footnote{The equations of motion of  General Relativity are independent
of $\xi_\mu$, but this is not necessarily true for more general actions.
We can decompose $\delta_L g_{\mu\nu}$ as $\delta_L g_{\mu\nu}
= h_{\mu\nu}+2\nabla_{(\mu} \xi_{\nu)}$ \cite{Battye:2013ida}.}, 
has both time-like parts and space-like parts 
\beq \label{eq:deformationvector}
        \xi_\mu = - \chi u_\mu + m_\mu.
\eeq
We can find the various contractions of $\delta_E {T^\mu}_\nu$ 
by inserting the derivatives of the deformation vector. 
We use two different
variational operators: $\delta_L$ and
$\delta_E$, which are linked via the Lie derivative along the diffeomorphism-generating
vector $\xi^{\mu}$ via
\beq
        \delta_L = \delta_E + L_{\xi},
\eeq
where $\delta_L$ is the Lagrangian perturbation (in co-moving coordinates)
and $\delta_E$ is the Eulerian perturbation (points are fixed in
spacetime). 
$L_\xi$ is the Lie derivative in the direction of $\xi^\mu$, a vector field
representing possible coordinate transformations, so that the Lie derivative acting on a given symmetric tensor field is $L_\xi X_{\mu\nu} = \xi^\alpha \nabla_\alpha
X_{\mu\nu}+2X_{\alpha(\mu} \nabla_\alpha \xi^{\nu)}$. Note that $L_\xi g_{\mu\nu} = 2\nabla_{(\mu} \xi_{\nu)}$. We are interested in the perturbed fluid equations, which are derived from
the Lagrangian for perturbations, denoted as $\ld_{(2)}$. 
In order to find the perturbed fluid variables, we first need to find 
the Eulerian change in the stress-energy tensor, $\delta_E {T^{\mu\nu}}$.
 We will work in the synchronous
gauge, 
where perturbations to the metric have spatial components only
\beq \label{eq:synch}
        \delta_E g_{\mu\nu} = {\gamma^\alpha}_\mu {\gamma^\beta}_\nu h_{\alpha\beta},
\eeq Working in the synchronous gauge (\ref{eq:synch}), we vary \eqref{eq:L} to obtain the 
contractions of the stress energy tensor \cite{Battye:2013ida,Battye:2013er}
\bse \label{eq:contractionsinpaper}
        \beq 
                u_\mu u^\nu \delta_E {T^\mu}_\nu = 
                & \left( \dot{\rho} + K (B_W + \rho) \right) \chi - (B _W+ \rho)
\left( \frac{1}{2} \hat{h} + \nabla_\alpha m^\alpha \right) 
                - (\rho + A_W) \dot{\chi}, \\ 
                u^\nu {\gamma^\sigma}_\mu \delta_E {T^\mu}_\nu = 
                &  (\frac{1}{4} C_W + P) \bar{\nabla}^\sigma \chi - \frac{1}{4}
C_W K^{\sigma\beta} m_\beta + (\frac{1}{4} C_W - \rho) \dot{m}^\sigma 
                + \rho {K^\sigma}_\alpha m^\alpha, \\ 
                {\gamma^\nu}_\mu \delta_E {T^\mu}_\nu = 
                & - \frac{1}{2} \left( \gamma^{\alpha\beta} ( 3D_W + E_W + P) + 3 u^\alpha u^\beta
                (B_W - P) \right) \delta_E g_{\alpha\beta} \\
                & - \left[ 3 u^\alpha u^\beta (B_W - P) + \gamma^{\alpha\beta} (3D _W+ E_W + P)
                \right] \nabla_\alpha \xi_\beta + 3 \chi \dot{P} 
                & \\
                {\gamma^\sigma}_\mu {\gamma^\nu}_\rho  \delta_E {T^\mu}_\nu
                = & \chi {\gamma^\sigma}_\rho \dot{P} - \frac{1}{2} \left(
(D_W + P){\gamma^\sigma}_\rho \hat{h} 
                + (E_W - 2 P) {\hat{h}^\sigma}_\rho       \right) \\ \nonumber
                & + {\gamma^\sigma}_\rho \left[ (P - A_W) \dot{\chi} - (D_W +
P) \left( \bar{\nabla}_\alpha m^\alpha - \chi K \right) \right] \\ 
                \nonumber
                & + (2P - F_W) \left[ \bar{\nabla}^{(\sigma} m_{\rho)} - m_\beta
K^{\beta( \sigma} u_{\rho)} - \chi {K^\sigma}_\rho \right], 
                \nonumber
        \eeq
\ese
where we have defined ``time" and ``space" differentiation as the derivative
operator projected along the time and space directions
\beq \label{eq:timeandspacedifferentiation}
        \dot{\psi} \equiv u^\mu \nabla_\mu \psi,  \hspace{35pt} \bar{\nabla}_\mu
\psi \equiv {\gamma^\nu}_\mu \nabla_\nu \psi,
\eeq
and where $K= 3 H$, (where $H$ is the Hubble parameter)
is the trace of the extrinsic curvature tensor
$K_{\mu\nu}\equiv \nabla_\mu u_\nu$, which satisfies $K_{\mu\nu}=K_{(\mu\nu)}$ and 
$u^\mu K_{\mu\nu}=0$.
\subsection{Perturbed fluid variables}
The components of the perturbed energy-momentum tensor ${T^\mu}_\nu $ are
written as
\beq \label{eq:fluidvar}
        \delta_E {T^\mu}_\nu = \delta \rho u^\mu u_\nu + 2 (\rho + P) v^{(\mu}
u_{\nu)} + \delta P {\gamma^\mu}_\nu + P {\Pi^\mu}_\nu,
\eeq
where $v^\mu$ is the perturbed dark sector velocity and $P{\Pi^\mu}_\nu$ is
the anisotropic part of the stress tensor, which is orthogonal and symmetric.
We have dropped the subscript $E$ on the variation of the density and pressure. 
We can now compare (\ref{eq:fluidvar}) and the contractions of the perturbed
stress energy tensor \eqref{eq:contractionsinpaper}, to obtain 
the perturbed fluid variables 
in terms of the deformation vector \eqref{eq:deformationvector},
as described in \cite{Battye:2013ida,Battye:2013er}
\bse \label{eq:per}
        \beq  
                \delta \rho = & \left(\dot{\rho} + K (B_W + \rho) \right) \chi-
(\rho + A_W) \dot{\chi}- (B_W + \rho) \left( \frac{1}{2} \hat{h} + \bar{\nabla}_\alpha
m^\alpha \right) 
                , \\
                \delta P = & (P - B_W) \dot{\chi} -  \frac{1}{3} ( 3D _W+ E_W +
P) \left( \frac{1}{2} \hat{h} + \bar{\nabla}_\alpha m^\alpha \right) \\ 
      \nonumber
                & + \left[ \frac{1}{3} ( 3D _W+ E_W + P)  K + \dot{P} \right]
\chi, \\ 
                (\rho + P) v^\sigma = & \left(\rho - \frac{C_W}{4} \right)
\dot{\bar{m}}^\sigma - \left( P + \frac{C_W}{4} \right) \bar{\nabla}^\sigma
             \chi,  \\
                P {\Pi^\sigma}_\rho = & (2P - E_W) \left[ \frac{1}{2} {\hat{h}^\sigma}_\rho
+ \bar{\nabla}^{(\sigma} m_{\rho)} 
                - \frac{1}{3} {\gamma^\sigma}_\rho \left( \frac{1}{2} \hat{h}
+ \bar{\nabla}^\alpha m_\alpha \right) \right], \\ \nonumber
        \eeq
\ese
where we have defined  $P {\Pi^\mu}_\nu = \left({\gamma^\mu}_\beta {\gamma^\alpha}_\nu
- \frac{1}{3} {\gamma^\alpha}_\beta {\gamma^\mu}_\nu \right)\delta_E
{T^\beta}_\alpha$.

\subsection{Invariance under time reparameterisation}
We now want to discover what constraints invariance under changes in time
imposes. This is helpful for \cite{Battye:2013er,Pearson:2014iaa} 
because applying time translational reparameterisation invariance but not
spatial reparameterisation invariance leads to elastic dark energy. 
Hence we set the coefficients of $\chi$, $\dot{\chi}$ and $\bar{\nabla}^\sigma
\chi$ in (\ref{eq:per}) to zero and therefore obtain
\beq \label{eq:densityevol}
        \dot{\rho} + 3H (P + \rho) &=& 0, \nonumber\\
        \dot{P} + \left( P + 3D _W+ E_W \right) H &=& 0. \label{eq:pressure}
\eeq
This gives the conservation equation and an equation for the evolution of
pressure with time. We can rewrite the perturbed fluid variables (\ref{eq:per})
as
\bse \label{eq:penulper}
        \beq 
                \delta \rho & = & - (P + \rho) \left( \frac{1}{2} \hat{h}
+ \bar{\nabla}_\alpha \xi^\alpha \right), \\ 
                \delta P & = & -  \beta \left( \frac{1}{2} \hat{h} + \bar{\nabla}_\alpha
\xi^\alpha \right), \\ 
                v^\sigma & = & \dot{\xi}^\sigma, \\ 
                P {\Pi^\sigma}_\rho & = & 2 \mu \left[ \frac{1}{2} {\hat{h}^\sigma}_\rho
+ \bar{\nabla}^{(\sigma} \xi_{\rho)} 
                - \frac{1}{3} {\gamma^\sigma}_\rho \left( \frac{1}{2} \hat{h}
+ \bar{\nabla}^\alpha \xi_\alpha \right) \right], 
        \eeq
\ese
and the pressure evolution
as
\beq \label{eq:betapressure} 
        \dot{P} + 3 \beta H = 0,
\eeq
where we have defined
\bse
        \beq 
                \beta &=& \frac{1}{3} P + D _W+ \frac{1}{3}E_W,  \label{eq:defofbeta} \\ 
                \mu &=& P - \frac{1}{2} E_W, \label{eq:defofmu} 
        \eeq
\ese
as parameters that can be determined by experiment. $\beta$ and $\mu$ correspond
to the elastic bulk modulus and the elastic shear modulus,
respectively \cite{Battye:2013er,Pearson:2014iaa}.

So far, we have summarised previous work. However, as an aside 
we note that it is possible to place constraints on the parameters using
observational data. In \cite{Battye:2014xna}, it was noted that
it might be possible to place constraints  on the parameters of the 
elastic dark energy model, referred to as Time Diffeomorphism Invariant (TDI) 
models in their paper. The authors used observational data on  cosmic shear and
CMB lensing which would give constraints for any given value of $w$, 
unless $w\approx-1$, as a wide . Unfortunately, recent Planck data
has shown that $w\approx -1$, so these constraint do not apply.

In Appendix \ref{sec:constraints}, using 2018 Planck data \cite{Ade:2015xua}, 
we find a constraint of $-0.004<\beta<0.106$ if we assume that the 
equation of state parameter $w$ of the Lagrangian \eqref{eq:L}, where $P=w\rho$, is constant.
If we assume that $w\neq -1$ exactly, then we also find constraints of 
$-0.0477\leq \hat{\mu} \leq 0.0599$, where we have defined $\hat{\mu}=\mu/\rho$. 
However,the assumption that we do not have a phantom equation of state, 
i.e. we require $w\geq -1$, gives $0\leq \hat{\mu} \leq 0.0599$.  

\newpage
\section{$\mathcal{L}_{\{2\}}(\delta_L u_\mu, \delta_L \gamma_{\mu\nu})$} 
\label{eq:secmetricsplit}
In the previous section, we summarised the derivation of the perturbed fluid variables for 
theories where the second variation of the Lagrangian depends only on the change
in the metric \cite{Battye:2013er} and found new
constraints on the values of the elasticity and rigidity parameters. We now move on 
to extend this work to more general Lagrangians.

If we take the second variation of the Lagrangian as a function only of 
the change in the time and spatial parts
of the metric separately
\beq \label{eq:moregenerallagrangiangeneric}
        \ld_{\{ 2 \} } = \ld_{\{2\}} \left(\delta_L u_\mu, \delta_L \gamma_{\mu\nu} \right), 
\eeq
then the most general possible quadratic Lagrangian
which is a function of $\delta_L u_\mu$ and $\delta_L \gamma_{\mu\nu}$ is 
\beq \label{eq:moregenerallagrangian}
        \bea
                \ld_{\{ 2 \} } = & \frac{1}{8} X^{\mu\nu\alpha\beta} \delta_L
\gamma_{\mu\nu} \delta_L \gamma_{\alpha\beta} 
                + \frac{1}{8} Y^{\mu\nu} \delta_L u_\mu \delta_L u_\nu +
\frac{1}{4} Q^{\mu\nu\alpha} \delta_L u_\mu \delta_L \gamma_{\nu\alpha},
        \eea
\eeq
where 
\beq \label{eq:symmofxyztensors}
        \bel
                X^{\mu\nu\alpha\beta} = X^{(\mu\nu)(\alpha\beta)} 
                = X^{\alpha\beta\mu\nu},\quad \quad
                Y^{\mu\nu} = Y^{(\mu\nu)}, \quad \quad
                Q^{\mu\nu\alpha} = Q^{\mu(\nu\alpha)}.
        \eea
\eeq
Using the identities \cite{Pearson:2014iaa}
\ba
         \delta_L u^\mu &=& \frac{1}{2} u^\mu u^\alpha u^\beta \delta_L g_{\alpha\beta},\non\\         \delta_L \gamma_{\mu\nu} &=& \delta_L g_{\mu\nu}
        + 2 u_{(\mu} \left(\gamma^\alpha{}_{\nu)} - \frac{1}{2} u_{\nu)}  u^\alpha \right) u^\beta \delta_L g_{\alpha\beta},
\ea         it is possible
to write \eqref{eq:moregenerallagrangian} as 
\ba \label{eq:variationofugammaitovarig}
        \ld_{\{ 2 \} } = \left(W_X^{\mu\nu\alpha\beta} + W_Y^{\mu\nu\alpha\beta}
        + W_Z^{\mu\nu\alpha\beta}\right) \delta_L g_{\alpha\beta} \delta_L g_{\mu\nu},
\ea      
where upon comparing \eqref{eq:moregenerallagrangian} with \eqref{eq:L}, we can see that the components of $W^{\sigma\rho\phi\lambda}=
W^{\sigma\rho\phi\lambda}_X + W^{\sigma\rho\phi\lambda}_Y + W^{\sigma\rho\phi\lambda}_Q$
are
\bse
        \beq
                W^{\sigma\rho\phi\lambda}_X &= & \frac{1}{2} X^{\mu\nu\alpha\beta}
\left[ 2 \left( {\delta^{(\sigma}}_\mu {\delta^{\rho)}}_\nu 
                {\delta^{(\phi}}_\alpha {\delta^{\lambda)}}_\beta + {\delta^{(\phi}}_\mu
{\delta^{\lambda)}}_\nu {\delta^{(\sigma}}_\alpha 
                {\delta^{\rho)}}_\beta \right) + 4 u_{(\mu} {H^{\sigma\rho}}_{\nu)}
\right. \\
                && \left. {\delta^{(\phi}}_\alpha {\delta^{\lambda)}}_\beta
+ 4 u_{(\alpha} {H^{\phi\lambda}}_{\beta)} {\delta^{(\sigma}}_\mu 
                {\delta^{\rho)}}_\nu + 4 u_{(\mu} {H^{\sigma\rho}}_{\nu)}
u_{(\alpha} {H^{\phi\lambda}}_{\beta)} \right],\\
        W^{\sigma\rho\phi\lambda}_Y &= & \frac{1}{4} Y^{\mu\nu} \left[ {H^{\phi\lambda}}_\mu
{H^{\sigma\rho}}_\nu + {H^{\phi\lambda}}_\nu 
        {H^{\sigma\rho}}_\mu \right], \\
        W^{\sigma\rho\phi\lambda}_Q &=& 2 Q^{\mu\nu\alpha} \left( {\delta^{(\sigma}}_\mu
{\delta^{\rho)}}_\nu + u_{(\mu} 
        {H^{\sigma\rho}}_{      \nu)} \right) {H^{\phi\lambda}}_\alpha,
\eeq
\ese
where we have defined
\beq
        {H^{\alpha\beta}}_\mu \equiv \left( \gamma^{(\alpha}_\mu - \frac{1}{2} u_\mu u^{(\alpha} \right) u^{\beta)}.
\eeq
\eqref{eq:variationofugammaitovarig} is a very important result -  
we can rewrite any Lagrangian dependent on the variation of the spatial part 
of the metric and the time-like part of the metric separately into
one dependent on only the variation of the full metric.

In the rest of this section, we will explore how we can use 
the method of section \ref{{sec:varfullmetriconly}} to find the 
perturbations of a Lagrangian of the form \eqref{sec:varfullmetriconly} without any new calculations. 
\subsection{Decomposition of the $X$,$Y$ and $Q$ tensors}
The most
general tensors that satisfy the necessary symmetries 
\eqref{eq:symmofxyztensors} are:
\bse
        \beq
                X^{\mu\nu\alpha\beta} &= &  A_X u^\mu
u^\nu u^\alpha u^\beta +B_X \left( u^\mu u^\nu \gamma^{\alpha\beta}
+ \gamma^{\mu\nu} u^\alpha u^\beta \right) 
                + 4C_X u^{(\mu} \gamma^{\nu)(\alpha} u^{\beta)} , \\ \nonumber
                && + D_X \gamma^{\mu\nu} \gamma^{\alpha\beta} + \frac{1}{2} E_X \gamma^{\mu(\alpha}
\gamma^{\beta)\nu}, \\ 
                Y^{\mu\nu} & =&  A_Y u^\mu u^\nu +B_Y \gamma^{\mu\nu} , \\
                Q^{\mu\nu\alpha} & =& A_Q u^\mu u^\nu u^\alpha +B_Q u^\mu \gamma^{\nu\alpha} + 2C_Q
u^{(\nu} \gamma^{\alpha)\mu}  . 
        \eeq
\ese
\subsection{The specific $W$ tensor for $\mathcal{L}(\delta_L u_\mu, \delta_L \gamma_{\mu\nu})$} 
Collecting like terms, we obtain 
\beq
                W^{\sigma\rho\phi\lambda} 
                &= & W^{\sigma\rho\phi\lambda}_X + W^{\sigma\rho\phi\lambda}_Y
+ W^{\sigma\rho\phi\lambda}_Q \non\\
                &= & \left( \frac{1}{2} A_X + \frac{1}{8} A_Y + \frac{1}{2}
A_Q \right) u^\sigma u^\rho u^\phi u^\lambda + B_X \left( u^\sigma u^\rho \gamma^{\phi\lambda} + \gamma^{\sigma\rho}
u^\phi u^\lambda \right) 
                 \non\\
                && + \left( 8C_X + \frac{1}{2} B_Y + B_Q + 2C_Q \right) u^{(\sigma}
\gamma^{\rho)(\phi} u^{\lambda)}+ 2 D_X \gamma^{\sigma\rho} \gamma^{\phi\lambda} + E_X
\gamma^{\phi(\sigma} \gamma^{\rho)\lambda}.
\eeq
Comparing to (\ref{eq:1}), we find 
\beq \label{eq:values}
                A_W &= & \frac{1}{2} A_X + \frac{1}{8} A_Y + \frac{1}{2}
A_Q, \nonumber\\
                B_W &= &  B_X , \nonumber\\
                C_W &= & 8C_X + \frac{1}{2} B_Y + B_Q + 2C_Q,\nonumber\\
                D_W &= & D_X, \nonumber\\ 
                E_W &= &  E_X. 
\eeq
\subsection{Invariance under time reparameterisation} 
We repeat our calculations from earlier, where we fixed our equations
to be the same under a change in time and we obtained (\ref{eq:betapressure}).

Plugging in the values from (\ref{eq:values}), we find an equation for the
evolution of the pressure
\beq
        \dot{P} = & - \left( P + 3D_X +E_X \right) H = & - 3 \beta_g H, 
\eeq
where $\beta_g\equiv \frac{1}{3} \left( P + 3D_X + E_X \right)$.
We also find that the coefficients of the $W$ tensor become 
\beq \label{eq:Lofugamma} 
         A_W & = \frac{1}{2} A_X + \frac{1}{8} A_Y + \frac{1}{2}
A_Q &= \rho, \nonumber\\
                B_W& = B_X&=P, \nonumber\\
                 C_W & =  8C_X + \frac{1}{2} B_Y + B_Q + 2C_Q &= -P, \nonumber\\
                 D_W & = D_X,& \nonumber\\ 
                 E_W & =  E_X.&  
\eeq
Next we will examine the specific cases where the Lagrangian depends only
on the variation of either the time \textit{or} the spatial part of the metric.
\subsubsection{$\ld_{\{2\}}(\delta_L u_\mu)$}

First, if we make $\ld_{\{2\}}$ a function of $\delta_L u_\mu$ only, 
i.e. $\ld_{\{2\}}$ is dependent only
on the change in the time part of the metric, we obtain $ W_X = W_Q = 0,$ in which case
$\beta=P= 0$, and using (\ref{eq:defofmu}) we can then rewrite the perturbed fluid variables
from (\ref{eq:penulper}) 
\bse
        \beq
                        \delta \rho & = - \rho \left( \frac{1}{2} \hat{h}
+ \bar{\nabla}_\alpha \xi^\alpha \right), \\
                        \delta P & = 0, \\ 
                        (1 + w) v^\sigma & = \dot{\xi}^\sigma, \\
                        P {\Pi^\sigma}_\rho & = 0.
        \eeq
\ese

We find that the equation of state parameter is 
\beq
        w=\dot{w}  = 0.
\eeq
Using constraints from Planck \cite{Ade:2015xua} which show $w\approx-1$, we can therefore rule out this case. 
While this case might intuitively seem unlikely,\footnote{The lack of spatial dependence means
that the dark energy Lagrangian is described by a pressureless fluid.} we have completely ruled it out without having to examine any
specific model. This shows the power of our parameterisation method.

\subsubsection{ $\ld_{\{2\}}(\delta_L \gamma_{\mu\nu})$}
If we make $\ld$ a function of $ \delta_L \gamma_{\mu\nu}$  only, i.e. $\ld$ is dependent
only on the change in the spatial part of the metric, we get $W_Y=W_Q=0$,
which means 
\beq \label{eq:gammavar}
        \begin{array}{rrl}
                A_W & =&  \frac{1}{2} A_X =\rho,\\
                  B_W & =&  B_X= P, \\
                C_W & =&  8 C_X=-P, \\
                 D_W & =& D_X, \\ 
                 E_W & =&  E_X, \\
        \eea
\eeq
and
\beq
        \bea
                \dot{P} = & - \left( P + 3D_X + E_X \right) H \\
                \equiv & -3\beta_\gamma H. 
        \eea
\eeq
where we have defined $\beta_\gamma=\frac{1}{3}\left(P + 3D_X + E_X \right)$.
For $\ld_{\{2\}}(\delta_L \gamma_{\mu\nu})$ the evolution equation for $w$ (\ref{eq:evoofw})
remains the same but with $\beta$ now depending on $D_X$ and $E_X$ rather than $D$ and
$F$, which lead to the same result if we use \eqref{eq:gammavar}.

\subsubsection{Summary of $\ld_{\{2\}}(\delta_L u_\mu, \delta_L \gamma_{\mu\nu})$}
The analysis of Section \ref{eq:secmetricsplit} shows that
the effect of the metric split is simply to change the coefficients as 
shown in \eqref{eq:values}.  

If we have mandated time reparameterisation invariance by decoupling
$\chi$, then the only relevant contribution to
$\beta$ and $\mu$ come from the spatial part of the metric $\gamma_{\mu\nu}$, i.e. only the first term
in \eqref{eq:moregenerallagrangian} has any effect on the system.

\subsection{Comparison with elastic dark energy theories}
We can compare to \cite{Pearson:2014iaa} to find the properties of the 
dark energy material, as this Lagrangian can be described using elastic dark energy.

We perform a scalar-vector-tensor (SVT) decomposition, where we decompose the perturbation to the metric
$h_{ij}$ as \cite{Battye:2007aa}
\ba \label{eq:svtpertdecom}
       \frac{1}{2} h_{ij}=  H^S_L Q^S_{ij} + H^S_T Q^S_{ij}
       +H^V Q^V_{ij} +H^T Q^T_{ij},
\ea
where one can decompose a spatial tensor field as
\ba
        \eta^{ij} \nabla_i \nabla_j Q^{S,V,T} = -k^2 Q^{S,V,T},
\ea        
scalars can be constructed from vectors and tensors as \cite{Hu:1997mn}
\ba
          \nabla_i Q^S = - k Q^S_i, \quad \quad
          \nabla_i \nabla_j Q^S + \frac{1}{3}k^2 \eta_{ij} Q^S = Q^S_{ij},         
\ea        
and vectors can be constructed from tensors as 
\ba
       \nabla_{(i} Q^V_{j)} = - k Q^V_{ij},
\ea         
with the requirement that $Q^{V|i}_i = Q^{T|i}_{ij} =Q^{Ti}_i = 0$.
We now find the entropy perturbation, which is defined by 
\ba
        w\Gamma = \left(\frac{\delta P}{\delta \rho} - \frac{dP}{d\rho}\right) \delta.      
\ea        
The entropy perturbation and scalar anisotropic stress are given by
\ba
        w \Gamma &=& 0, \non\\
        w \Pi^S &=& -2 \frac{\mu}{\rho+P} \left[ \delta -3(1+w)\eta\right],
\ea
where we have defined $\delta=\frac{\delta \rho}{\rho}$, 
and $\eta=-\left(H^S_L +\frac{1}{3}H^S_T\right)$.
This is the same result as \cite{Pearson:2014iaa} finds when looking at 
elastic mediums.

\newpage
\section{Tensor-metric theories}
So far, we have looked at theories where $\ld_{\{2\}} = \ld_{\{2\}}(\delta g_{\mu\nu}, \delta f_{\mu\nu})$. 
In this section, we now apply our analysis to theories where the dark Lagrangian is a function of both the variation of the
 metric $g_{\mu\nu}$ and of an unspecified non-dynamical symmetric 
 rank-2 tensor $f_{\mu\nu}$ \cite{deRham:2010kj,Schmidt-May:2015vnx} \footnote{It should be noted that 
at this point there are no derivatives of $f_{\mu\nu}$ in the action and therefore
no kinetic term for $f_{\mu\nu}$. We could choose a form of $f_{\mu\nu}$
that includes derivatives of a vector field or a scalar, and therefore generates a kinetic term.}. 

If $\ld_{\{2\}} = \ld_{\{2\}}(\delta g_{\mu\nu}, \delta f_{\mu\nu})$, then the most general quadratic Lagrangian 
we can have is
\beq \label{eq:bigenerallagrangian}
\ld_{\{2\}} = \frac{1}{8} A^{\mu\nu\alpha\beta} \delta_L g_{\mu\nu} \delta_L g_{\alpha\beta} + \frac{1}{4} B^{\mu\nu\alpha\beta} 
\delta_L g_{\mu\nu} \delta_L f_{\alpha\beta} + \frac{1}{8} C^{\mu\nu\alpha\beta} \delta_L f_{\mu\nu} \delta_L f_{\alpha\beta}.
\eeq
The tensors obey the following symmetries
\beq
                A^{\mu\nu\alpha\beta} = A^{(\mu\nu)(\alpha\beta)} = A^{\alpha\beta\mu\nu}, \quad
                \quad 
                B^{\mu\nu\alpha\beta} = B^{(\mu\nu)(\alpha\beta)}, \quad
                \quad 
                C^{\mu\nu\alpha\beta} = C^{(\alpha\beta)(\mu\nu)}.
\eeq
The decomposition of these tensors, called ``coupling tensors'' because they prescribe how the fields combine in the Lagrangian, is 
\bse \label{eq:decompofbicouplingtensors}
        \beq \label{eq:decompofa}
                A^{\mu\nu\alpha\beta} &=&   A_X u^{\mu} u^{\nu} u^{\alpha} u^{\beta} + B_X \left( u^{\mu} u^{\nu} \gamma^{\alpha\beta} 
                + \gamma^{\mu\nu} u^{\alpha} u^{\beta} \right) 
                + 4C_X u^{(\mu} \gamma^{\nu)(\alpha} u^{\beta)} \nonumber \\
                && + D_X \gamma^{\mu\nu} \gamma^{\alpha\beta} 
                + 2 E_X \gamma^{\mu(\alpha} \gamma^{\beta)\nu},\\       
                B^{\mu\nu\alpha\beta}&= & A_Y u^{\mu} u^{\nu} u^{\alpha} u^{\beta}+ B_Y  u^{\mu} u^{\nu} \gamma^{\alpha\beta} + 4 C_Y u^{(\mu} \gamma^{\nu)(\alpha} u^{\beta)} 
                 \nonumber\\
                && + D_Y \gamma^{\mu\nu} \gamma^{\alpha\beta} + 2 E_Y \gamma^{\mu(\alpha} \gamma^{\beta)\nu} 
                + F_Y \gamma^{\mu\nu} u^{\alpha} u^{\beta},\\
                C^{\mu\nu\alpha\beta}& = &A_Z u^{\mu} u^{\nu} u^{\alpha} u^{\beta}+ B_Z \left( u^{\mu} u^{\nu} \gamma^{\alpha\beta} + \gamma^{\mu\nu} u^{\alpha} u^{\beta} \right) 
                + 4 C_Z u^{(\mu} \gamma^{\nu)(\alpha} u^{\beta)}   \nonumber\\
                && + D_Z \gamma^{\mu\nu} \gamma^{\alpha\beta} + 2 E_Z \gamma^{\mu(\alpha} \gamma^{\beta)\nu}.
        \eeq
\ese

\subsection{Equations of motion}
We use \eqref{eq:decompofbicouplingtensors} to find 
\beq
        \bea
                \delta_E {T^\mu}_\nu = & - \frac{1}{2} \left( B_X u^\mu u_\nu + D_X  {\gamma^\mu}_\nu + {T^\mu}_\nu \right) h - E_X {h^\mu}_\nu 
                - \frac{1}{2} {{B^\mu}_\nu}^{\alpha\beta} k_{\alpha\beta} \\
                & - \left( \nabla_\alpha {T^\mu}_\nu - \frac{1}{2} {{B^\mu}_\nu}^{\sigma\beta} \nabla_\alpha f_{\sigma\beta} \right) \xi^\alpha \\
                & + \left( 2 T^{\alpha(\mu} {g^\beta}_{\nu)} - {{B^\mu}_\nu}^{\sigma\alpha} {f^\beta}_\sigma - {{A^\mu}_\nu}^{\alpha\beta} 
                - {T^\mu}_\nu g^{\alpha\beta} \right) \nabla_\alpha \xi_\beta 
        \eea
\eeq
where we have defined $k_{\mu\nu} \equiv \delta_E f_{\mu\nu}$.
\newpage
We find the perturbed fluid variables 
by assuming the unperturbed tensor $f_{\mu\nu}$ is homogenous, i.e. 
$\bar{\nabla}_\alpha f_{\mu\nu}=0$, as chosen in \cite{Lagos:2016gep}, 
recalling the definition of ``space'' differentiation 
$\bar{\nabla}_\mu
\psi \equiv {\gamma^\nu}_\mu \nabla_\nu \psi$ from
\eqref{eq:timeandspacedifferentiation}
\bse \label{eq:biperv}
        \beq
                \delta \rho &=& - (B_X + \rho) \left( \frac{1}{2}h +\bar{\nabla}^\alpha m_\alpha \right) 
                - \frac{1}{2} \left( B_Y \gamma^{\alpha\beta} + A_Y u^\alpha u^\beta \right) k_{\alpha\beta}\nonumber \\
                && + \left[ \dot{\rho} + \frac{1}{2} \dot{f}^{\alpha\beta} \left( B_Y \gamma_{\alpha\beta} + A_Y u_\alpha u_\beta \right)  
                + B_Y f^{\alpha\beta} K_{\alpha\beta} + ( \rho + B_X) K \right] \chi \nonumber\\
                && - (A_X + A_Y {f^\beta}_\sigma u^\sigma u_\beta + \rho) \dot{\chi} - f^{\alpha\beta} B_Y \bar{\nabla}_\alpha m_\beta, \\
                (\rho + P) v^\lambda &=& (\rho - C_X) \dot{\bar{m}}^\lambda - (P + C_X -  C_Y u^\sigma u_\beta {f^\beta}_\sigma) 
                \bar{\nabla}^\lambda \chi \nonumber\\
                && -C_Y \bigg\{ \gamma^{\lambda(\alpha} u^{\beta)} k_{\alpha\beta} 
                + \gamma^{\lambda\sigma} {f^\beta}_\sigma \dot{m}_\beta - m_                \beta f^{\beta\sigma} {K^\lambda}_\sigma \bigg\}, \\ \nonumber
                \delta P &= & - \frac{1}{3} \left( 3 D_X + P + 2 E_X \right) \left( \frac{1}{2} h 
                + \bar{\nabla}^\beta m_\beta \right) - \left(\frac{1}{2} B_Y u^\alpha u^\beta + \frac{1}{6}(3 D_Y + 2 E_Y)\gamma^{\alpha\beta} \right) k_{\alpha\beta}\nonumber \\
                && - f^{\alpha\beta} \frac{1}{3} (3 D_Y + 2 E_Y) \bar{\nabla}_\alpha m_\beta 
                + (P - B_X + B_Y f^{\alpha\beta} u_\alpha u_\beta) \dot{\chi} \nonumber \\
                && +  \bigg[ \dot{P} + \frac{1}{3} \left( P + 3 D_X + 2 E_X \right) K + \frac{1}{3} (3 D_Y + 2 E_Y) f^{\alpha\beta} K_{\alpha\beta}    \nonumber\\
                & &- \frac{1}{2} \dot{f}^{\alpha\beta} \left( \frac{1}{3} \left( 3 D_Y + 2 E_Y \right) \gamma_{\alpha\beta} 
                + A_Y u_\alpha u_\beta \right) \bigg] \chi, \\ \nonumber 
                P {\Pi^\rho}_\lambda &= & 2 (P - 2E_X) \left[ \frac{1}{2} {h^\rho}_\lambda + \bar{\nabla}^{(\rho} m_{\lambda)} 
                -\frac{1}{3} \gamma^\rho{}_\lambda \left(\frac{1}{2} h + \bar{\nabla}_\alpha m^\alpha\right)
                  \right] \non\\
                  &&- E_Y \bigg[ \left(\gamma^{\rho(\alpha} {\gamma^{\beta)}}_\lambda
                   -\frac{1}{3}\gamma^\rho{}_\lambda \gamma^{\alpha\beta}\right)
                 k_{\alpha\beta}+ 2 f^{\alpha\beta} \left(
                 {\gamma_\alpha}^{(\rho} \bar{\nabla}_{\lambda)} m_\beta 
                  -   -\frac{1}{3}\gamma^\rho{}_\lambda \bar{\nabla}_\alpha m_\beta\right) \bigg].~~~
                \eeq
\ese

Assuming time reparameterisation invariance gives
\bse \label{eq:nonlitimeperv}
        \beq 
                \delta \rho &= & - (A_X + \rho) \left( \frac{1}{2} h + \bar{\nabla}^\alpha \xi_\alpha \right) 
                - \frac{1}{2} \left( B_Y \gamma^{\alpha\beta} + A_Y u^\alpha u^\beta \right) 
                k_{\alpha\beta}
                 - B_Y f^{\alpha\beta}\bar{\nabla}_\alpha \xi_\beta, \\ 
                (\rho + P) v^\lambda &= & (\rho - C_X) \dot{\xi}^\lambda 
                - C_Y \left[\gamma^{\lambda(\alpha} u^{\beta)} k_{\alpha\beta} 
                + f^{\alpha\beta} \left( {\gamma^\lambda}_\alpha \dot{\xi}_\beta 
                -  {K^\lambda}_\alpha \xi_\beta  \right) \right] , \\ 
                \delta P &= & - \frac{1}{3} \left( 3 D_X + P + 2 E_X \right) \left( \frac{1}{2} h 
                + \bar{\nabla}^\beta \xi_\beta \right) 
                 - \left(\frac{1}{2} B_Y u^\alpha u^\beta + \frac{1}{6}(3 D_Y + 2 E_Y)
                 \gamma^{\alpha\beta} \right) k_{\alpha\beta}~~~\nonumber \\
                && - \frac{1}{3} f^{\alpha\beta} (3 D_Y + 2 E_Y) \bar{\nabla}_\alpha \xi_\beta, \\ \nonumber  
                P {\Pi^\rho}_\lambda &= & 2 (P - 2E_X) \left[ \frac{1}{2} {h^\rho}_\lambda 
                + \bar{\nabla}^{(\rho} \xi_{\lambda)} 
                -\frac{1}{3} \gamma^\rho{}_\lambda \left(\frac{1}{2} h + \bar{\nabla}_\alpha \xi^\alpha\right)
                  \right] \non\\
                  &&- E_Y \bigg\{ \left(\gamma^{\rho(\alpha} {\gamma^{\beta)}}_\lambda
                   -\frac{1}{3}\gamma^\rho{}_\lambda \gamma^{\alpha\beta}\right)
                 k_{\alpha\beta}+ 2 f^{\alpha\beta} \left(
                 {\gamma_\alpha}^{(\rho} \bar{\nabla}_{\lambda)} \xi_\beta 
                     -\frac{1}{3}\gamma^\rho{}_\lambda \bar{\nabla}_\alpha \xi_\beta\right) \bigg\},
        \eeq
\ese
and we obtain evolution equations for 
$P$ and $\rho$
\beq \label{eq:nlievo}
        \bea \label{eq:tensormetrictheoriesevolutioneqns}
                \dot{\rho} = & - \frac{1}{2} \dot{f}^{\alpha\beta} \left( B_Y \gamma_{\alpha\beta} + A_Y u_\alpha u_\beta \right) - B_Y f^{\alpha\beta} K_{\alpha\beta} - (\rho + B_X) K, \\
    \dot{P} = & \frac{1}{3} (3 D_Y + 2 E_Y) f^{\alpha\beta} K_{\alpha\beta} + \frac{1}{2} (D_Y \gamma_{\alpha\beta} - B_Y u_\alpha u_\beta) 
                \dot{f}^{    \alpha\beta}  - \frac{1}{3} ( P + 3 D_X + 2 E_X) K. 
        \eea
\eeq
\section{No preferred direction in the coupling tensors}
\subsection{Preferred direction}
Using $u_\mu$\ and $\gamma_{\mu\nu}$ in the decomposition of the coupling tensors means that we have chosen a preferred direction for the Lagrangian. 
If we assume there is no preferred direction and only the full metric is seen in the coupling tensors,
then \eqref{eq:decompofbicouplingtensors} becomes
\bse
        \beq \label{eq:Lorentzinv}
                A^{\mu\nu\alpha\beta} = A_A g^{\mu\nu} g^{\alpha\beta} + 2 B_A g^{\alpha(\mu} g^{\nu)\beta}, \\ 
                B^{\mu\nu\alpha\beta} = A_B g^{\mu\nu} g^{\alpha\beta} + 2 B_B g^{\alpha(\mu} g^{\nu)\beta}, \\
                C^{\mu\nu\alpha\beta} = A_C g^{\mu\nu} g^{\alpha\beta} + 2 B_C g^{\alpha(\mu} g^{\nu)\beta}. 
        \eeq
\ese
In order to obtain (\ref{eq:Lorentzinv}) we must set the coefficients in \eqref{eq:decompofbicouplingtensors} as follows 
\beq \label{eq:licomponentsofabc}
        \begin{array} {rlrlrl}
                A_X &= A_A + 2 B_A & A_Y & =  A_B + 2 B_B & A_Z & =A_C + 2 B_C \\
                B_X &= - A_A & B_Y & = - A_B & B_Z & =  - A_C \\
                C_X &= - B_A & C_Y & =- B_B & C_Z & = - B_C \\
                D_X &= A_A   & D_Y & =   A_B & D_Z & =   A_C \\
                E_X &= B_A   & E_Y & =   B_B & E_Z & =   B_C\\
                                &        & F_Y & = - A_B. &     &    \\
        \eea
\eeq

\subsection{The perturbed fluid variables}
\label{sec:bilichange}
When we use the coupling tensors (\ref{eq:licomponentsofabc}), 
then we find that the perturbed fluid variables when $\chi$ is decoupled are
\bse \label{eq:bilitimeperv}
        \beq
                \delta \rho &= &  (A_A - \rho) \left(\frac{1}{2} h + \bar{\nabla}_\alpha \xi^\alpha \right) 
                + A_B \left( f^{\alpha\beta} \nabla_\alpha \xi_\beta + \frac{1}{2} k \right) 
                - \frac{1}{3} B_B k^{\alpha\beta} u_\alpha u_\beta, \\ 
                (\rho + P) v^\sigma &= & B_B \left[u^\nu {\gamma^\sigma}_\mu {k^\mu}_\nu + B_B \left( {\gamma^\sigma}_\mu f^{\mu\beta}
                \dot{\xi}_\beta - f^{\alpha\beta} {K^\sigma}_\alpha \xi_\beta \right) \right] + (B_A + \rho) \dot{\xi}^\sigma, \\ 
                \delta P &= & - \left( \frac{1}{3} \left(2 B_A + P + 3 A_A \right) \left( \frac{1}{2} h + \bar{\nabla}_\alpha \xi^\alpha \right) 
                + \frac{1}{2} A_B k +  \frac{1}{3} B_B {\gamma^\nu}_\mu {k^\mu}_\nu \right) \\ \nonumber
                && - \frac{1}{3} \left( 3 A_B  + 2 B_B \right) f^{\alpha\beta} \bar{\nabla}_\alpha \xi_\beta, \\ 
                P {\Pi^\rho}_\lambda &= & 2 (P - 2B_A) \left( \frac{1}{2} {h^\rho}_\lambda 
                + \bar{\nabla}^{(\rho} \xi_{\lambda)}-\frac{1}{3} \gamma^\rho{}_\lambda \left(\frac{1}{2} h 
                + \bar{\nabla}_\alpha \xi^\alpha\right)  \right)\non\\
                && -2 B_B\left[  {\gamma^\rho}_\mu {\gamma^\nu}_\lambda {k^\mu}_\nu  -\frac{1}{3}\gamma^\nu{}_\mu
                 k^\mu{}_\nu \gamma^\rho{}_\lambda               
                + 2 f^{\alpha\beta}\left( \gamma_{\alpha(\lambda} \bar{\nabla}^{\rho)} \xi_\beta 
                - \frac{1}{3} \gamma^\rho{}_\lambda \bar{\nabla}_\alpha \xi_\beta
                 \right)\right], 
        \eeq
\ese
and we can write the evolution equations \eqref{eq:tensormetrictheoriesevolutioneqns} as 
\bse
        \beq
                \dot{\rho} = & A_B (f^{\alpha\beta} K_{\alpha\beta} + \frac{1}{2} \dot{f}) - B_B \dot{f}^{\alpha\beta} u_\alpha u_\beta 
                + (A_A - \rho) K \\
                \dot{P} = & - \frac{1}{6}  \bigg( 2( P + 3 A_A + 2 B_A ) K + 3 A_B \left( \dot{f} 
                + 2 f^{\alpha\beta} K_{\alpha\beta} \right)  + 2 B_B 
                \left(  \dot{f}^{\alpha\beta} \gamma_{\alpha\beta} + 2 f^{\alpha\beta} K_{\alpha\beta} \right) \bigg). 
        \eeq
\ese

\newpage
\section{Choosing a form for $f_{\mu\nu}$}
\subsubsection{Conformal and disformal choices}
An obvious choice for $f_{\mu\nu}$ is
\ba
        f_{\mu\nu} = A_f \phi g_{\mu\nu} + B_f \nabla_\mu \phi \nabla_\nu \phi
\ea        
where $A_f$ and $B_f$ are constants and $\phi$ is a scalar field.
However, because this means that the variation of $f_{\mu\nu}$ contains the variation of the metric,
we must modify \eqref{eq:bigenerallagrangian}. In fact, this choice means that \eqref{eq:bigenerallagrangian}
can be rewritten as the second variation of a Lagrangian of the form $\ld_{(2)}(\delta_OL g_{\mu\nu},
\delta_L \phi,  \D_\mu \delta_L \phi)$, i.e. 
\ba \label{eq:secvarsclaar}
        \mathcal{L}_{(2)} &=& \mathcal{A}(\delta_L \phi)^2
        + \mathcal{B}^\mu \delta_L \phi \D_\mu \delta_L \phi
        + \frac{1}{2} \mathcal{C}^{\mu\nu} \D_\mu \delta_L
        \phi \D_\nu \delta_L \phi \non\\
        && +\frac{1}{4}\left[ \mathcal{Y}^{\alpha\mu\nu}
        \D_\alpha \delta_L \phi \delta_L g_{\mu\nu}
        + \mathcal{V}^{\mu\nu} \delta_L \phi \delta_L g_{\mu\nu}
        +\frac{1}{2} \mathcal{W}^{\mu\nu\alpha\beta}
        \delta_L g_{\mu\nu} \delta_L g_{\alpha\beta}\right],        
\ea        
which was studied in \cite{Battye:2012eu}. The terms 
on the second line of \eqref{eq:secvarsclaar} could be possibly be generated by ``Beyond Horndeski'' 
theories \cite{Gleyzes:2014dya,BenAchour:2016fzp}. 
The Beyond Horndeski theories contain terms in a specific combination 
to avoid the Ostrogradski instability, which would place constraints on the 
couplings $A$, $B^\mu$ and $C^{\mu\nu}$.

\subsubsection{Flat reference metric}
Another choice is a flat reference metric, i.e. $f_{\mu\nu}=\eta_{\mu\nu}$, 
although this choice does not simplify our perturbed fluid variables greatly
while we are still using the generalised lagrangian \eqref{eq:bigenerallagrangian}.

\section{Summary of results}
In this paper, we have 
\begin{itemize}
        \item
        summarised the calculations given in \cite{Battye:2013ida} for the perturbed fluid variables for a general dark
        Lagrangian of the form $\ld_{\{2\}}  = \ld_{\{2\}}(\delta_L g_{\mu\nu})$ by working in 
        the synchronous gauge and in the perfectly elastic case,
        both with and without the imposition of time reparameterisation invariance        
        \item
        obtained new constraints on the values of $w$ and $\mu$ for a
        realistic sound speed under various conditions, using data from the Planck satellite
        \item
        rewritten the perturbed fluid variables for general dark
        Lagrangians of the form\\ $\ld_{\{2\}}  = \ld_{\{2\}}(\delta_L \gamma_{\mu\nu}, \delta_L u_\mu)$, 
        $\ld_{\{2\}}  = \ld_{\{2\}}(\delta_L u_\mu)$ and $\ld_{\{2\}}  = \ld_{\{2\}}(\delta_L \gamma_{\mu\nu})$ both in
        general and when time reparameterisation invariance is imposed
        \item
                obtained new evolution equations for $\dot{P}$ and $w$ for
these new Lagrangians and repeated these calculations for tensor-metric theories where
$\ld _{\{2\}}= \ld_{\{2\}}(\delta_L g_{\mu\nu}, \delta_L f_{\mu\nu})$, using either a time-spatial 
metric split in the coupling tensors or using only the full metric 
        \item
                found new evolution equations for $\rho$ and $P$ and rewritten
the perturbed fluid variables in these theories         
\end{itemize}

There are many different modified gravity theories which attempt to
explain the accelerated expansion of the universe. 
By parameterising theories based on the second variation of the Lagrangian,
we were able to develop a framework which can very quickly rule
out various theories, and indeed we can rule out any theory which is purely a function
 of the variation of the time part of the metric.  Our method could be used both in 
 future models of massive gravity or to rule out large classes
of theories when new observational results are found.

We have examined the connections between these theories and elastic dark energy. 
For our first case, the elastic dark energy framework can be straightforwardly used. We also placed constraints on the parameters of elastic dark energy using Planck results.
Future work could examine whether other dark energy models can use this framework.

\section{Acknowledgements}
We would like to thank Richard Battye and David Burton for valuable discussions and suggestions, 
both on the content of this paper and the presentation.
\bibliographystyle{unsrt}
\bibliography{bibliography}

\begin{thebibliography}{10}

\bibitem{Perlmutter:1998np}
S.~Perlmutter et~al.
\newblock {Measurements of Omega and Lambda from 42 high redshift supernovae}.
\newblock {\em Astrophys. J.}, 517:565--586, 1999.

\bibitem{Riess:1998cb}
Adam~G. Riess et~al.
\newblock {Observational evidence from supernovae for an accelerating universe
  and a cosmological constant}.
\newblock {\em Astron. J.}, 116:1009--1038, 1998.

\bibitem{Heymans:2012gg}
Catherine Heymans et~al.
\newblock {CFHTLenS: The Canada-France-Hawaii Telescope Lensing Survey}.
\newblock {\em Mon. Not. Roy. Astron. Soc.}, 427:146, 2012.

\bibitem{Ade:2013kta}
P.~A.~R. Ade et~al.
\newblock {Planck 2013 results. XV. CMB power spectra and likelihood}.
\newblock {\em Astron. Astrophys.}, 571:A15, 2014.

\bibitem{Ade:2013tyw}
P.~A.~R. Ade et~al.
\newblock {Planck 2013 results. XVII. Gravitational lensing by large-scale
  structure}.
\newblock {\em Astron. Astrophys.}, 571:A17, 2014.

\bibitem{Abbott:2005bi}
T.~Abbott et~al.
\newblock {The dark energy survey}.
\newblock 2005.

\bibitem{Abate:2012za}
Alexandra Abate et~al.
\newblock {Large Synoptic Survey Telescope: Dark Energy Science Collaboration}.
\newblock 2012.

\bibitem{Battye:2018ssx}
Richard~A. Battye, Francesco Pace, and Damien Trinh.
\newblock {Gravitational wave constraints on dark sector models}.
\newblock 2018.

\bibitem{Clifton:2011jh}
Timothy Clifton, Pedro~G. Ferreira, Antonio Padilla, and Constantinos Skordis.
\newblock {Modified Gravity and Cosmology}.
\newblock {\em Phys. Rept.}, 513:1--189, 2012.

\bibitem{Pearson:2012kb}
Jonathan~A. Pearson.
\newblock {Effective field theory for perturbations in dark energy and modified
  gravity}.
\newblock 2012.

\bibitem{Bloomfield:2013cyf}
Jolyon Bloomfield and Jonathon Pearson.
\newblock {Simple implementation of general dark energy models}.
\newblock {\em JCAP}, 1403:017, 2014.

\bibitem{Battye:2013er}
Richard~A. Battye and Jonathan~A. Pearson.
\newblock {Massive gravity, the elasticity of space-time and perturbations in
  the dark sector}.
\newblock {\em Phys.Rev.}, D88(8):084004, 2013.

\bibitem{Battye:2012eu}
Richard~A. Battye and Jonathan~A. Pearson.
\newblock {Effective action approach to cosmological perturbations in dark
  energy and modified gravity}.
\newblock {\em JCAP}, 1207:019, 2012.

\bibitem{Battye:2013ida}
Richard~A. Battye and Jonathan~A. Pearson.
\newblock {Computing model independent perturbations in dark energy and
  modified gravity}.
\newblock {\em JCAP}, 1403:051, 2014.

\bibitem{Skordis:2008vt}
Constantinos Skordis.
\newblock {Consistent cosmological modifications to the Einstein equations}.
\newblock {\em Phys. Rev.}, D79:123527, 2009.

\bibitem{Baker:2011jy}
Tessa Baker, Pedro~G. Ferreira, Constantinos Skordis, and Joe Zuntz.
\newblock {Towards a fully consistent parameterization of modified gravity}.
\newblock {\em Phys. Rev.}, D84:124018, 2011.

\bibitem{Baker:2012zs}
Tessa Baker, Pedro~G. Ferreira, and Constantinos Skordis.
\newblock {The Parameterized Post-Friedmann framework for theories of modified
  gravity: concepts, formalism and examples}.
\newblock {\em Phys. Rev.}, D87(2):024015, 2013.

\bibitem{TheLIGOScientific:2017qsa}
B. P. Abbott et~al.
\newblock {GW170817: Observation of Gravitational Waves from a Binary Neutron
  Star Inspiral}.
\newblock {\em Phys. Rev. Lett.}, 119(16):161101, 2017.

\bibitem{deRham:2018red}
Claudia de~Rham and Scott Melville.
\newblock {Gravitational Rainbows: LIGO and Dark Energy at its Cutoff}.
\newblock 2018.

\bibitem{Pearson:2014iaa}
Jonathan~A. Pearson.
\newblock {Material models of dark energy}.
\newblock {\em Annalen Phys.}, 526:318--339, 2014.

\bibitem{Carter:1982xm}
B.~Carter.
\newblock {Interaction of gravitational waves with an elastic solid medium}.
\newblock In {\em {Les Houches Summer School on Gravitational Radiation Les
  Houches, France, June 2-21, 1982}}, 1982.

\bibitem{Carter:1973zz}
B.~Carter.
\newblock {Speed of Sound in a High-Pressure General-Relativistic Solid}.
\newblock {\em Phys. Rev.}, D7:1590--1593, 1973.

\bibitem{Carter:1980zz}
B.~Carter.
\newblock Rheometric structure theory, convective differentiation and continuum
  electrodynamics.
\newblock {\em Proceedings of the Royal Society of London. Series A,
  Mathematical and Physical Sciences}, 372(1749):169--200, 1980.

\bibitem{Friedman:1975ab}
J.~L. {Friedman} and B.~F. {Schutz}.
\newblock {On the stability of relativistic systems}.
\newblock {\em The Astrophysical Journal}, 200:204--220, August 1975.

\bibitem{Carter:1972az}
B.~Carter and H.~Quintana.
\newblock Foundations of general relativistic high-pressure elasticity theory.
\newblock {\em Proceedings of the Royal Society of London. Series A,
  Mathematical and Physical Sciences}, 331(1584):57--83, 1972.

\bibitem{Tattersall:2017eav}
Oliver~J. Tattersall, Macarena Lagos, and Pedro~G. Ferreira.
\newblock {Covariant approach to parametrized cosmological perturbations}.
\newblock {\em Phys. Rev.}, D96(6):064011, 2017.

\bibitem{Battye:2014xna}
Richard~A. Battye, Adam Moss, and Jonathan~A. Pearson.
\newblock {Constraining dark sector perturbations I: cosmic shear and CMB
  lensing}.
\newblock {\em JCAP}, 1504:048, 2015.

\bibitem{Ade:2015xua}
P.~A.~R. Ade et~al.
\newblock {Planck 2015 results. XIII. Cosmological parameters}.
\newblock {\em Astron. Astrophys.}, 594:A13, 2016.

\bibitem{Battye:2007aa}
Richard~A. Battye and Adam Moss.
\newblock {Cosmological Perturbations in Elastic Dark Energy Models}.
\newblock {\em Phys. Rev.}, D76:023005, 2007.

\bibitem{Hu:1997mn}
Wayne Hu, Uros Seljak, Martin~J. White, and Matias Zaldarriaga.
\newblock {A complete treatment of CMB anisotropies in a FRW universe}.
\newblock {\em Phys. Rev.}, D57:3290--3301, 1998.

\bibitem{deRham:2010kj}
Claudia de~Rham, Gregory Gabadadze, and Andrew~J. Tolley.
\newblock {Resummation of Massive Gravity}.
\newblock {\em Phys. Rev. Lett.}, 106:231101, 2011.

\bibitem{Schmidt-May:2015vnx}
Angnis Schmidt-May and Mikael von Strauss.
\newblock {Recent developments in bimetric theory}.
\newblock {\em J. Phys.}, A49(18):183001, 2016.

\bibitem{Lagos:2016gep}
Macarena Lagos and Pedro~G. Ferreira.
\newblock {A general theory of linear cosmological perturbations: bimetric
  theories}.
\newblock {\em JCAP}, 1701(01):047, 2017.

\bibitem{Gleyzes:2014dya}
Jérôme Gleyzes, David Langlois, Federico Piazza, and Filippo Vernizzi.
\newblock {Healthy theories beyond Horndeski}.
\newblock {\em Phys. Rev. Lett.}, 114(21):211101, 2015.

\bibitem{BenAchour:2016fzp}
Jibril Ben~Achour, Marco Crisostomi, Kazuya Koyama, David Langlois, Karim Noui,
  and Gianmassimo Tasinato.
\newblock {Degenerate higher order scalar-tensor theories beyond Horndeski up
  to cubic order}.
\newblock {\em JHEP}, 12:100, 2016.

\end{thebibliography}
\newpage
\appendix

\section{Evolution of $w$}\label{sec:constraints}
We want to find an evolution equation for $w$, the equation of state parameter
where $P = w \rho$. Using the conservation equation, (\ref{eq:densityevol}) and (\ref{eq:betapressure}),
and defining 
\beq
        \hat{\beta} \equiv \frac{\beta}{\rho},
\eeq        
we obtain 
\beq \label{eq:evoofw}
\dot{w} = 3 \left[ w( 1 + w) - \hat{\beta} \right] H,
\eeq 
which notably does not depend on $\mu$.
Hence $w$ is constant if 
\beq \label{eq:betaforconstw}
\hat{\beta} = w (1 + w),
\eeq
and so 
\beq \label{eq:constw}
        w = -\frac{1}{2} \pm \frac{1}{2} \sqrt{1 + 4\hat{\beta}}, 
\eeq
gives a stable universe. Using 2018 Planck data \cite{Ade:2015xua}, 
the 68\% constraint on $w$ is $w=-1.028\pm0.032$ which in turn
gives a constraint of $-0.004<\beta<0.106$.
\subsection{Sound speed}
The sound speed for elastic dark energy is given by \cite{Pearson:2014iaa}
\beq \label{eq:speedofsound}
        {c_s}^2 \equiv \frac{\hat{\beta} + \frac{4}{3} \hat{\mu} }{1 + w},
\eeq
where $\hat{\mu}=\frac{\mu}{\rho}$. Using (\ref{eq:betaforconstw}), and in order that the sound speed fulfills $0 \leq {c_s}^2 \leq 1$, i.e. is real and sub-luminal, then
for constant $w$ we must have
\beq \label{eq:subluminalcs}
        - \frac{3}{4} w (1 + w) \leq \hat{\mu} \leq \frac{3}{4} ( 1 - w^2).
\eeq
If we set $\hat{\mu} = 0= \dot{w}$, then \eqref{eq:subluminalcs}
gives that either $w=-1$ exactly, or 
\beq
        0 \leq w \leq 1,
\eeq
which leads to a contradiction with the acceleration of the universe, as acceleration
requires $w<-\frac{1}{3}$. 
This means that either $w=-1$ or \emph{a stable universe with zero shear modulus cannot support acceleration
of the universe} and we therefore require a non-zero $\mu$. 
Using the 2018 Planck data \cite{Ade:2015xua} together with \eqref{eq:subluminalcs}
gives us constraints of $-0.0477\leq \hat{\mu} \leq 0.0599$. 
However,the assumption that we do not have a phantom equation of state, 
i.e. we require $w\geq -1$, gives $0\leq \hat{\mu} \leq 0.0599$.

\end{document}